\documentclass[12pt]{article}
\usepackage[dvips]{graphicx}
\usepackage{amsmath}
\usepackage{graphicx}
\usepackage{amsfonts}
\usepackage{amssymb}
\usepackage{epsfig}
\usepackage{subfigure}

\newcommand{\nn}{\nonumber}
\begin{document}
\title{\bf Reflection and transmission resonances and accuracy of the WKB method}

\author{T. Ngampitipan$^{a}$\hspace{1mm} and P. Boonserm$^{b}$\\
$^{a}$ {\small {\em Department of Physics, Chulalongkorn University,}}\\
{\small {\em Bangkok 10330, Thailand}}\\
$^{b}$ {\small {\em Department of Mathematics and Computer Science, Chulalongkorn University,}}\\
{\small {\em Bangkok 10330, Thailand}}
}

\date{30-31 May 2012; \LaTeX-ed \today}

\maketitle

\begin{abstract}
In this paper we calculate the transmission and reflection amplitudes of wave functions for different potentials such as the delta function, the rectangular barrier, the Eckart potential, and the Hulthen potential. We describe the relationship between these amplitudes and compute the reflection resonances between each potential. We describe the transmission and reflection probabilities using the WKB formula and compare the results with ones obtained from matching the boundary conditions. Furthermore, we use a two by two transfer matrix to calculate a rigorous bound on the transmission and reflection probabilities.
\vspace{5mm}


\noindent Presented at the 2nd Regional Conference on Applied and Engineering Mathematics (Penang, Malaysia); 30-31 May 2012

\end{abstract}

\section{Introduction}
One-dimensional quantum problems widely appear in any textbook on quantum mechanics \cite{Landshoff, Landau, Baym, Gasiorowicz, Capri, Stehle, Schiff, Cohen, Galindo, Park, Fromhold, Scharff, Messiah, Merzbacher, Singh, Mathews}. Although they are simple in mathematics and clear in physics, they lead to important and new results \cite{1D, Bogoliubov, grey, mil, Analytic, Shabat, phd}. In physics, there are a number of applications of one-dimensional quantum problems. For example, in acoustics, one might be interested in the propagation of sound waves down a long pipe, while in electromagnetism, one might be interested in the physics of wave-guides \cite{phd}.

In this paper, we study one-dimensional quantum problems for four selected potentials, the delta function, the rectangular barrier, the Eckart potential, and the Hulthen potential \cite{Ushveridze}. We will calculate exact and approximate transmission and reflection probabilities by matching boundary conditions and by using the WKB method respectively, and compare them to each other. We will also use a $2 \times 2$ transfer matrix to obtain a rigorous bound on these probabilities. Moreover, we will derive resonances of transmission and reflection probabilities.

\section{Conventions}
We are interested in solving the time-independent Schr\"{o}dinger equation \cite{Landshoff, Landau, Baym, Gasiorowicz, Capri, Stehle, Schiff, Cohen, Galindo, Park, Fromhold, Scharff, Messiah, Merzbacher, Singh, Mathews},
\begin{equation}
\left[-\frac{\hbar^{2}}{2m}\frac{\text{d}^{2}}{\text{d}x^{2}} + V(x)\right]\psi(x) = E\psi(x).
\end{equation}
We concentrate on a potential which is asymptotically constant, that is $V(x) \rightarrow V_{\pm\infty}$ as $x \rightarrow \pm\infty$. For $E > V(x)$ for all $x$, the asymptotic wave functions are given by
\begin{eqnarray}
\psi(x) \rightarrow \left\{
         \begin{array}{lll}
           t\dfrac{\exp\left(-ik_{+\infty}x\right)}{\sqrt{k_{+\infty}}}, & \text{if} & x \rightarrow +\infty \\
           \dfrac{\exp\left(-ik_{-\infty}x\right)}{\sqrt{k_{-\infty}}} + r\dfrac{\exp\left(ik_{-\infty}x\right)}{\sqrt{k_{-\infty}}}, & \text{if} & x \rightarrow -\infty \\
         \end{array}
       \right.,
\end{eqnarray}
where wave numbers are defined by
\begin{equation}
k_{\pm\infty}^{2} = \frac{2m\left(E - V_{\pm\infty}\right)}{\hbar^{2}}.
\end{equation}
We identify $t$ and $r$ as transmission and reflection amplitudes respectively. Therefore, the conservation of flux leads to the condition
\begin{equation}
T + R = 1,\label{conservation}
\end{equation}
where $T \equiv |t|^{2}$ and $R \equiv |r|^{2}$ are transmission and reflection probabilities respectively. A position at which a transmission or reflection probability is unity, is thus referred to as the occurrence of resonance.

Moreover, we can obtain a rigorous bound on the transmission probabilities by using a $2 \times 2$ transfer matrix. It gives \cite{Bogoliubov}
\begin{equation}
T_{\text{transmission}} \geq \text{sech}^{2}\left(\frac{1}{2}\int_{x_{1}}^{x_{2}}\left|k_{0} - \frac{k^{2}(x)}{k_{0}}\right|\text{d}x\right),\label{T2x2}
\end{equation}
where
\begin{equation}
k^{2}(x) = \frac{2m\left[E - V(x)\right]}{\hbar^{2}}.
\end{equation}
and $k(x) \rightarrow k_{0}$ if $x$ is outside the interval $(x_{1}, x_{2})$.

On the other hand, for $E < V(x)$ in the range $x_{1} < x < x_{2}$ we can obtain the transmission probability for any potential by using a useful technique called the WKB method. The result is given by \cite{Solid}
\begin{equation}
T_{w} = \exp\left[-2\sqrt{\frac{2m}{\hbar^{2}}}\int_{x_{1}}^{x_{2}}\sqrt{V(x) - E}\text{d}x\right].\label{Tw}
\end{equation}

\section{One-dimensional problems}\label{one dim prob}
In this section, we review main results in some one-dimensional problems. Four types of potentials are selected to discuss in this paper.

\subsection{A delta function potential}
\begin{figure}[pb]
\centerline{\psfig{file=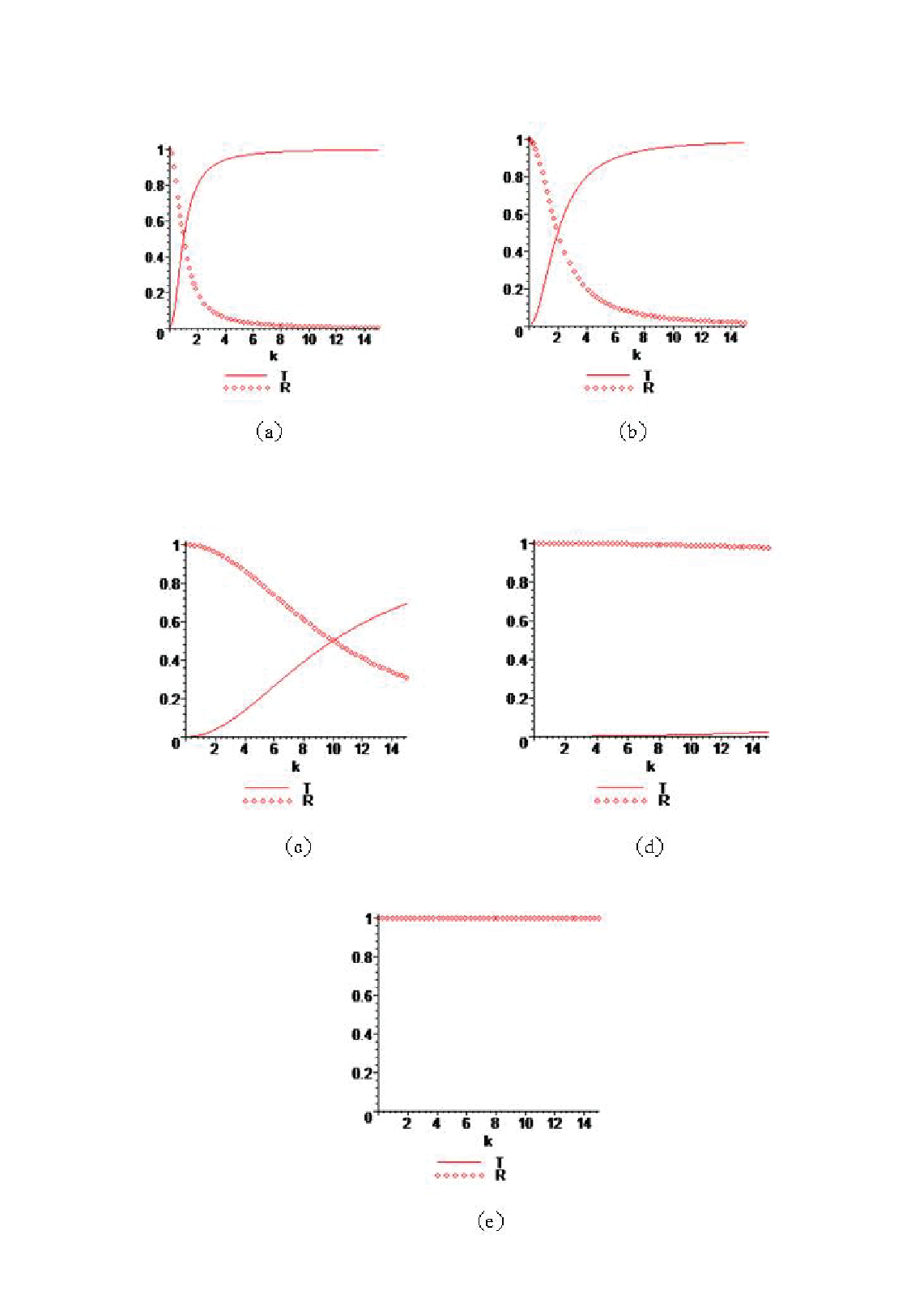,width=5in}}
\vspace*{1pt}
\caption{Plotting of transmission and reflection probabilities varying as $k$ for the delta function potential with (a) $k_{0} = 1$, (b) $k_{0} = 2$, (c) $k_{0} = 10$, (d) $k_{0} = 100$ and (e) $k_{0} = 1000$.}\label{Tdel}
\end{figure}

A delta function potential takes the form
\begin{equation}
V(x) = \alpha\delta(x),
\end{equation}
where $\alpha$ is a positive constant and the delta function is defined by
\begin{eqnarray}
\delta(x) = \left\{
         \begin{array}{lll}
           0,      & \text{if} & x \neq 0 \\
           \infty, & \text{if} & x = 0 \\
         \end{array}
       \right..
\end{eqnarray}
We are interested in a scattering state $E > 0$. We define
\begin{equation}
k^{2} = \frac{2mE}{\hbar^{2}}~~\text{and}~~k_{0} = \frac{m\alpha}{\hbar^{2}}.
\end{equation}
The transmission and reflection amplitudes for this potential are given by (see \cite{Baym, Gasiorowicz})
\begin{equation}
t = \frac{k}{k - ik_{0}}~~\text{and}~~r = \frac{ik_{0}}{k - ik_{0}}.
\end{equation}
Therefore, the transmission and reflection probabilities are
\begin{equation}
T = \frac{k^{2}}{k^{2} + k_{0}^{2}}~~\text{and}~~R = \frac{k_{0}^{2}}{k^{2} + k_{0}^{2}}.
\end{equation}
Note that
\begin{equation}
T + R = 1.
\end{equation}
The transmission and reflection probabilities varying with $k$ are shown in Fig. \ref{Tdel}. We have seen that the transmission probability tends to unify as $k$ goes to infinity. We say, however, that this potential has no transmission resonances. On the other hand, the reflection resonances occur at $k = 0$. Moreover, from the plotting if the potential is strongest, the total reflection happens. That is when the potential is stronger, penetration of a particle or wave through the potential is more difficult.

\subsection{Rectangular barrier potential}
\begin{figure}[pb]
\centerline{\psfig{file=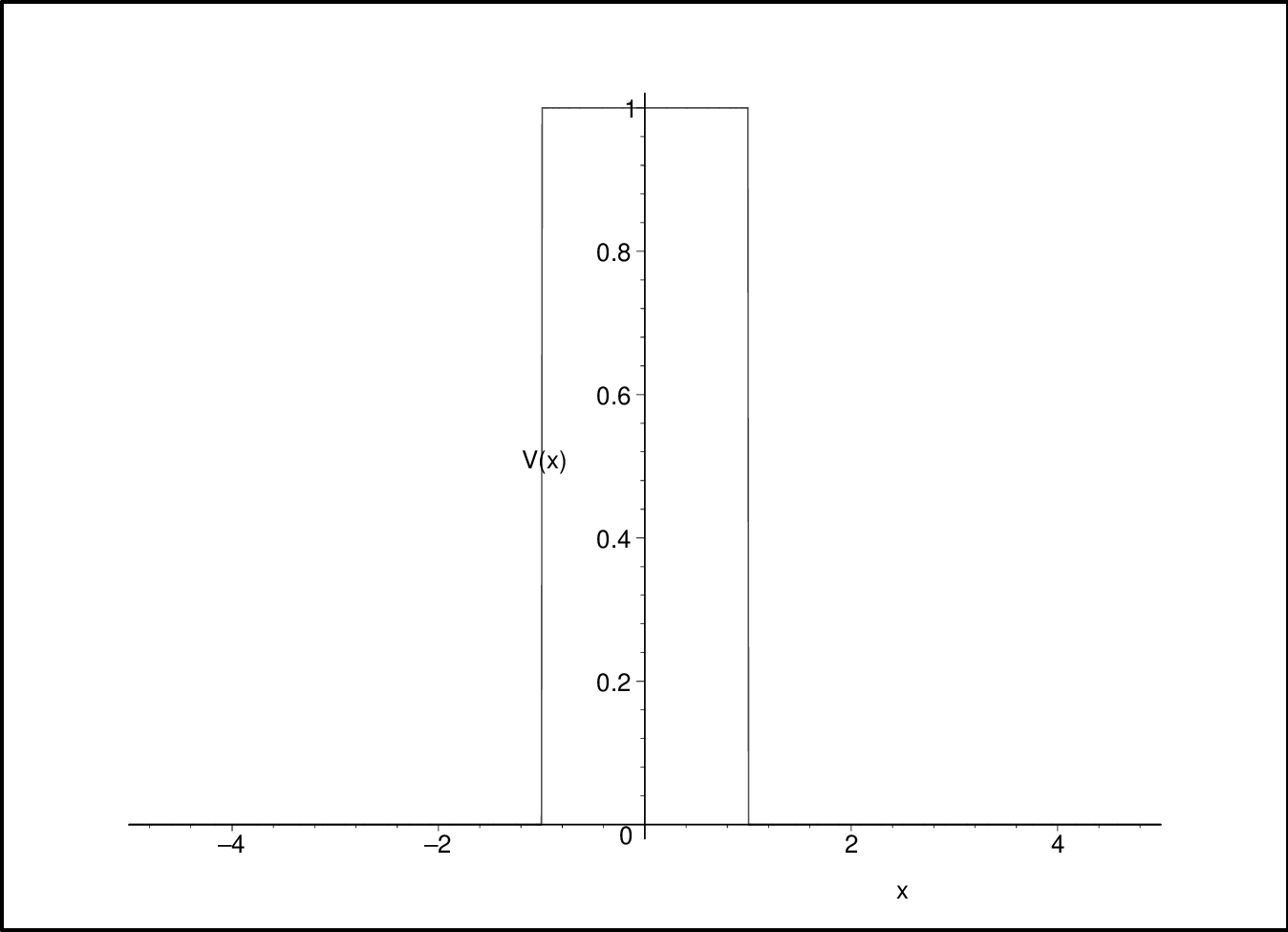,width=3in, angle = -90}}
\vspace*{1pt}
\caption{The rectangular barrier with $a = 1$ and $V_{0} = 1$.}\label{recpot}
\end{figure}

The rectangular barrier potential has the form (see \cite{Landau, Schiff})
\begin{eqnarray}
V(x) = \left\{
         \begin{array}{lll}
           V_{0}, & \text{if} & |x| \leq a \\
           0,     & \text{if} & \text{otherwise} \\
         \end{array}
       \right..
\end{eqnarray}
The shape of the rectangular barrier is shown in Fig. \ref{recpot}. We are interested in two cases $E > V_{0} > 0$ and $V_{0} > E > 0$.

\subsubsection{Case I: $E > V_{0} > 0$}
We define
\begin{equation}
k^{2} = \frac{2mE}{\hbar^{2}},~~q^{2} = \frac{2m(E - V_{0})}{\hbar^{2}},~~\text{and}~~k_{0}^{2} = \frac{2mV_{0}}{\hbar^{2}} = k^{2} - q^{2}.
\end{equation}
The transmission and reflection amplitudes for this potential are given by (see \cite{Messiah, Brandsen})
\begin{equation}
t = \frac{4kq\exp(2ika)}{(k + q)^{2}\exp(2iqa) - (k - q)^{2}\exp(-2iqa)}
\end{equation}
and
\begin{equation}
r = \frac{2i(k^{2} - q^{2})\sin(2qa)\exp(2ika)}{(k + q)^{2}\exp(2iqa) - (k - q)^{2}\exp(-2iqa)}.
\end{equation}
Therefore, the transmission and reflection probabilities are
\begin{equation}
T = \frac{4k^{2}q^{2}}{4k^{2}q^{2} + k_{0}^{4}\sin^{2}(2qa)}~~\text{and}~~R = \frac{k_{0}^{4}\sin^{2}(2qa)}{4k^{2}q^{2} + k_{0}^{4}\sin^{2}(2qa)}.
\end{equation}
Note that
\begin{equation}
T + R = 1.
\end{equation}

\begin{figure}[pb]
\centerline{\psfig{file=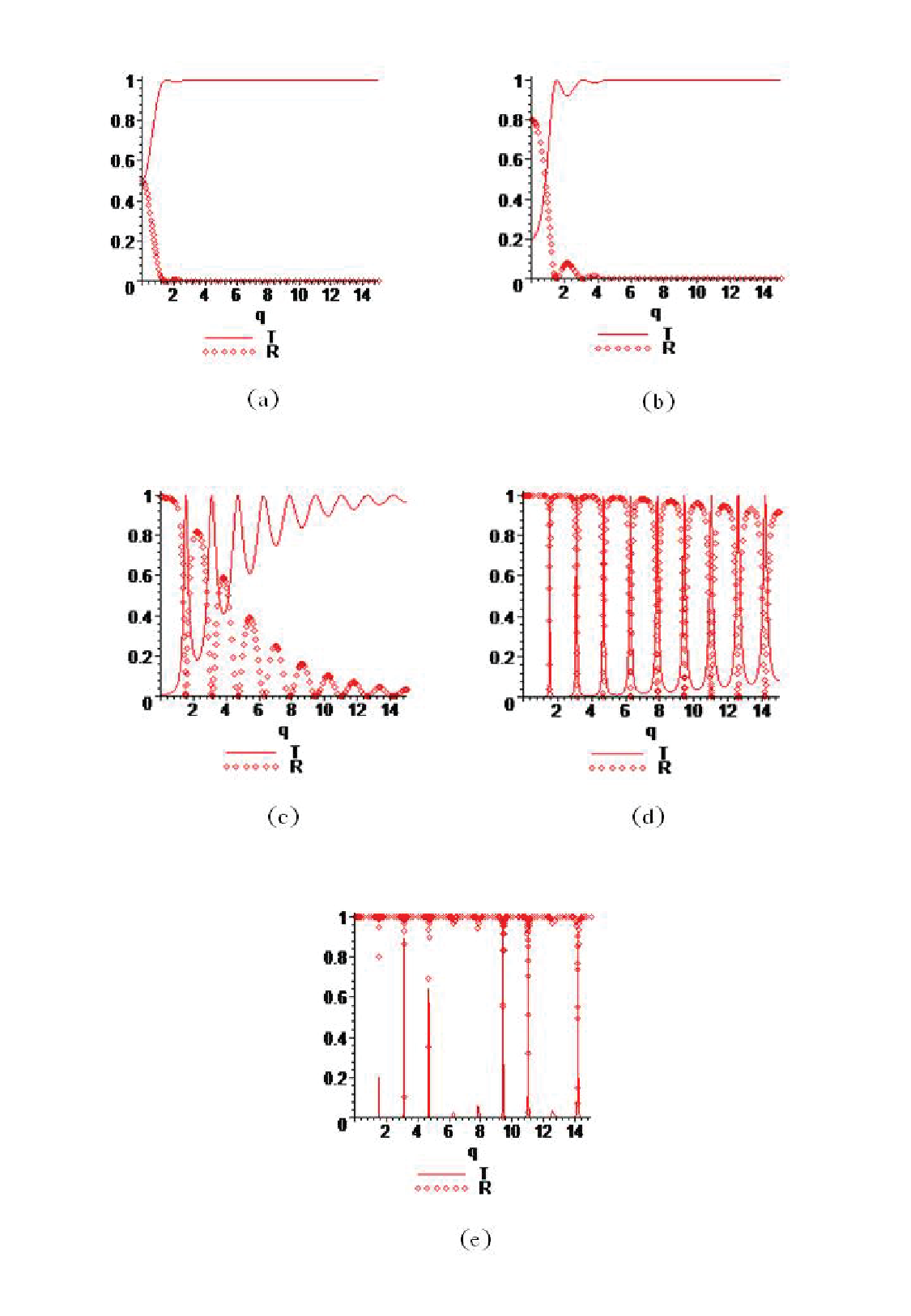,width=5in}}
\vspace*{1pt}
\caption{The effects of barrier heights $V_{0}$ of the potential on the probabilities for varying $q$ in the rectangular barrier with (a) $k_{0} = 1$, (b) $k_{0} = 2$, (c) $k_{0} = 10$, (d) $k_{0} = 100$ and (e) $k_{0} = 1000$ for $a = 1$.}\label{Trec}
\end{figure}

\begin{figure}[pb]
\centerline{\psfig{file=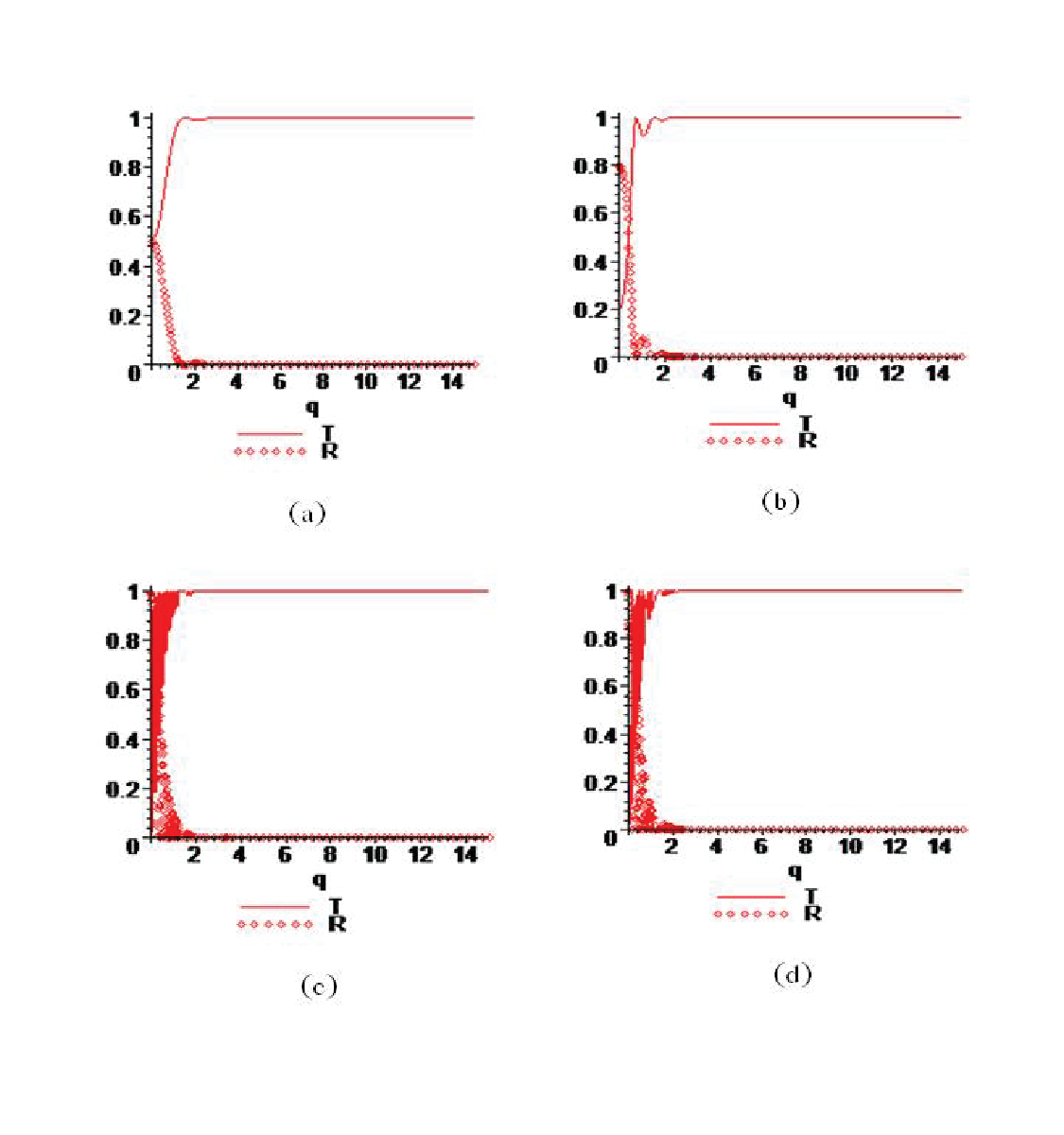,width=6in}}
\vspace*{1pt}
\caption{The effects of the widths $a$ of the potential on the probabilities varying as $q$ in the rectangular barrier with (a) $a = 1$, (b) $a = 2$, (c) $a = 10$, and (d) $a = 100$ for $k_{0} = 1$.}\label{Treca}
\end{figure}

The transmission and reflection probabilities varying with $q$ are shown in Fig. \ref{Trec} and \ref{Treca}. Fig. \ref{Trec} compares the effects of barrier heights $V_{0}$ of the potential on the probabilities when the width $a$ of the potential is fixed. It has been found that the higher the barrier of the potential is, the more the number of reflection resonances. This is similar to the case of the delta function potential: penetration of a particle or wave through the potential is hard to occur when the barrier height of the potential is large. Fig. \ref{Treca} compares the effects of the widths of the potential on the probabilities when the barrier height of the potential is fixed. The results are that the reflection resonance can occur when the width increases. Analytically, the transmission resonances occur at
\begin{equation}
q = \frac{n\pi}{2a},
\end{equation}
where $n = 1, 2, 3, ...$, while the reflection resonance is at $k = 0$ for this potential.

\begin{figure}[pb]
\centerline{\psfig{file=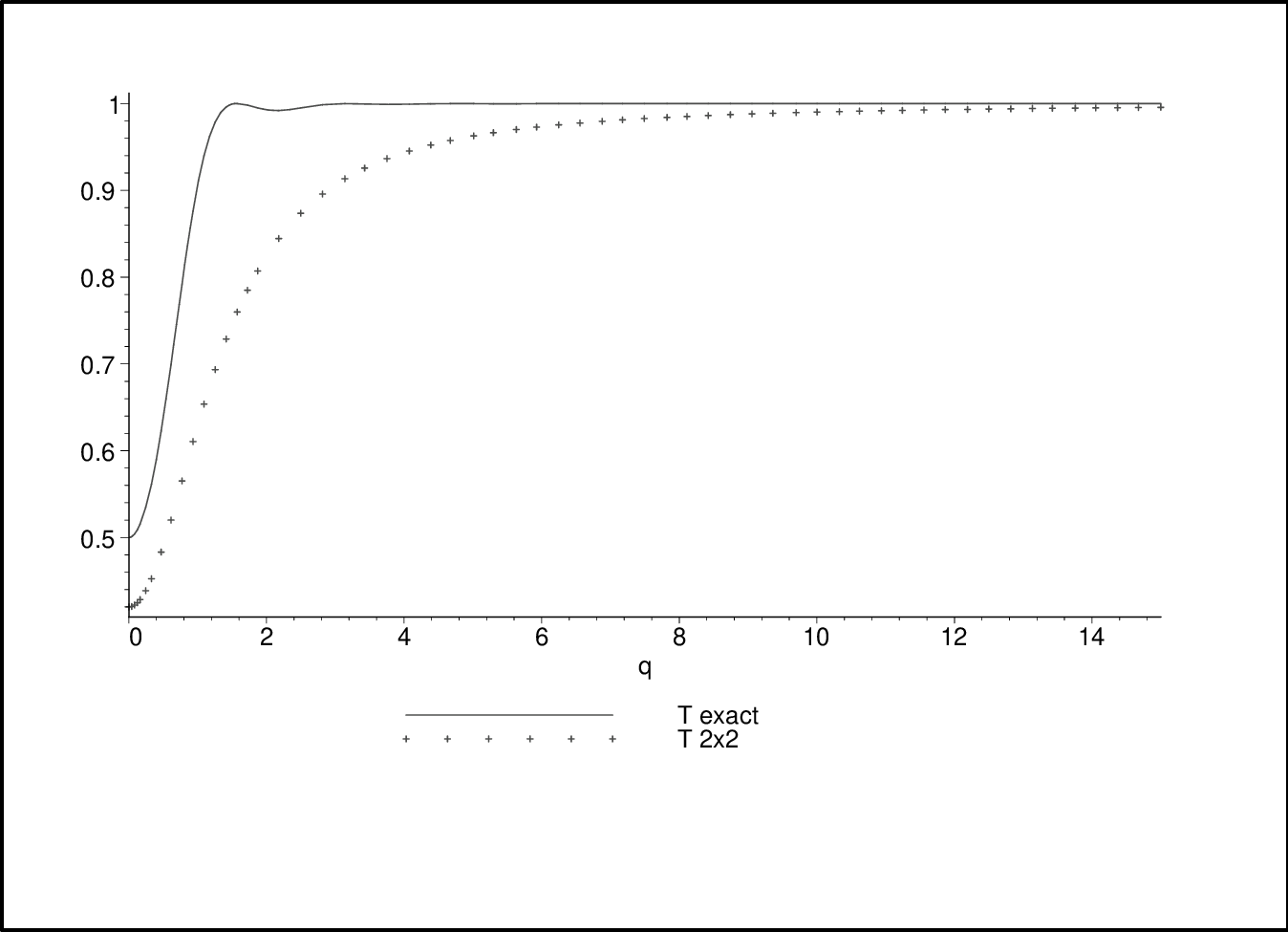,width=3in, angle = -90}}
\vspace*{1pt}
\caption{Plot of the transmission probability and its lower bound. $T$ $2 \times 2$ denotes the lower bound derived from a $2 \times 2$ transfer matrix.}\label{comprec}
\end{figure}

Moreover, we can obtain a lower bound of the transmission probability by using a $2 \times 2$ transfer matrix from equation (\ref{T2x2})
\begin{equation}
T \geq \text{sech}^{2}\left(\frac{k_{0}^{2}a}{\sqrt{k_{0}^{2} + q^{2}}}\right).
\end{equation}
This rigorous bound is compared with the exact solution in Fig. \ref{comprec}. It is found that the lower bound approaches the exact solution when a particle is at high energy.

\subsubsection{Case II: $V_{0} > E > 0$}
We define
\begin{equation}
Q^{2} = \frac{2m\left(V_{0} - E\right)}{\hbar^{2}}.
\end{equation}
The transmission and reflection amplitudes for this potential are given by
\begin{equation}
t = \frac{2iQke^{-2ika}}{\left(k^{2} - Q^{2}\right)\sinh(2Qa) + 2ikQ\cosh(2Qa)}
\end{equation}
and
\begin{equation}
r = \frac{\left(k^{2} + Q^{2}\right)\sinh(2Qa)e^{-2ika}}{\left(k^{2} - Q^{2}\right)\sinh(2Qa) + 2ikQ\cosh(2Qa)}.
\end{equation}
Therefore, the transmission and reflection probabilities are
\begin{equation}
T = \frac{4k^{2}Q^{2}}{\left(k^{2} - Q^{2}\right)^{2}\sinh^{2}(2Qa) + 4k^{2}Q^{2}\cosh^{2}(2Qa)}
\end{equation}
and
\begin{equation}
R = \frac{k_{0}^{4}\sinh^{2}(2Qa)}{\left(k^{2} - Q^{2}\right)^{2}\sinh^{2}(2Qa) + 4k^{2}Q^{2}\cosh^{2}(2Qa)}.
\end{equation}
We have seen that in this case $T \neq 0$. It follows that a particle can penetrate the barrier from one side to the other although the potential energy of the particle exceeds its total energy, which does not appear in classical physics. This is called tunneling. Note that in this case
\begin{equation}
T + R = 1.
\end{equation}

\begin{figure}[pb]
\centerline{\psfig{file=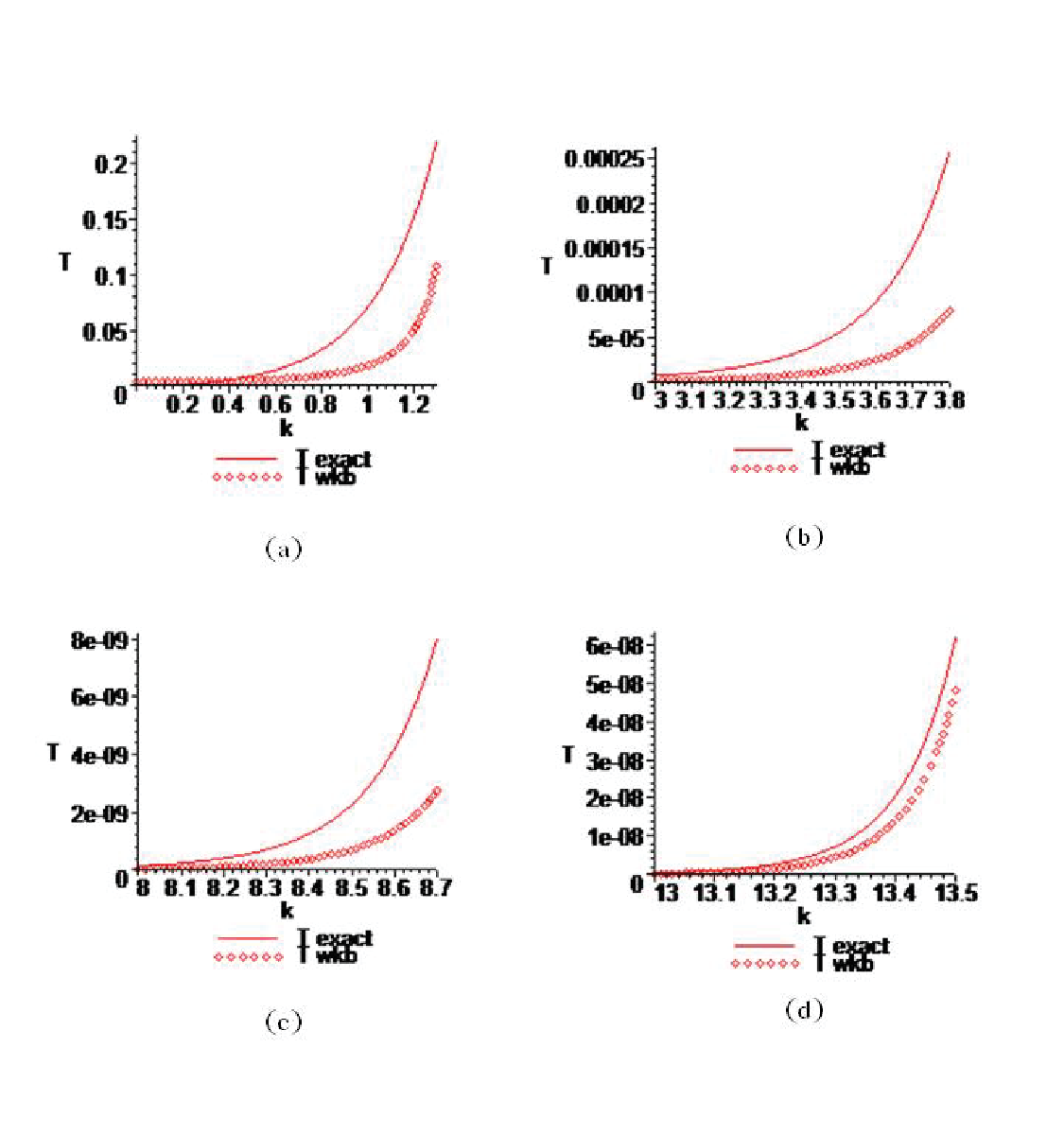,width=6in}}
\vspace*{1pt}
\caption{The exact solution of the transmission probability compared to the WKB approximation with (a) $V_{0} = 1$, (b) $V_{0} = 10$, (c) $V_{0} = 50$, and (d) $V_{0} = 100$ for $m = \hbar = a = 1$.}\label{Trectun}
\end{figure}

By the WKB method, we obtain
\begin{equation}
T_{w} = \exp(-4Qa).
\end{equation}
In Fig. \ref{Trectun}, the result from the WKB method is compared with the exact result. It is found that the larger the height of the potential, the more accurate the WKB approximation becomes.

\subsection{An Eckart potential}
\begin{figure}[pb]
\centerline{\psfig{file=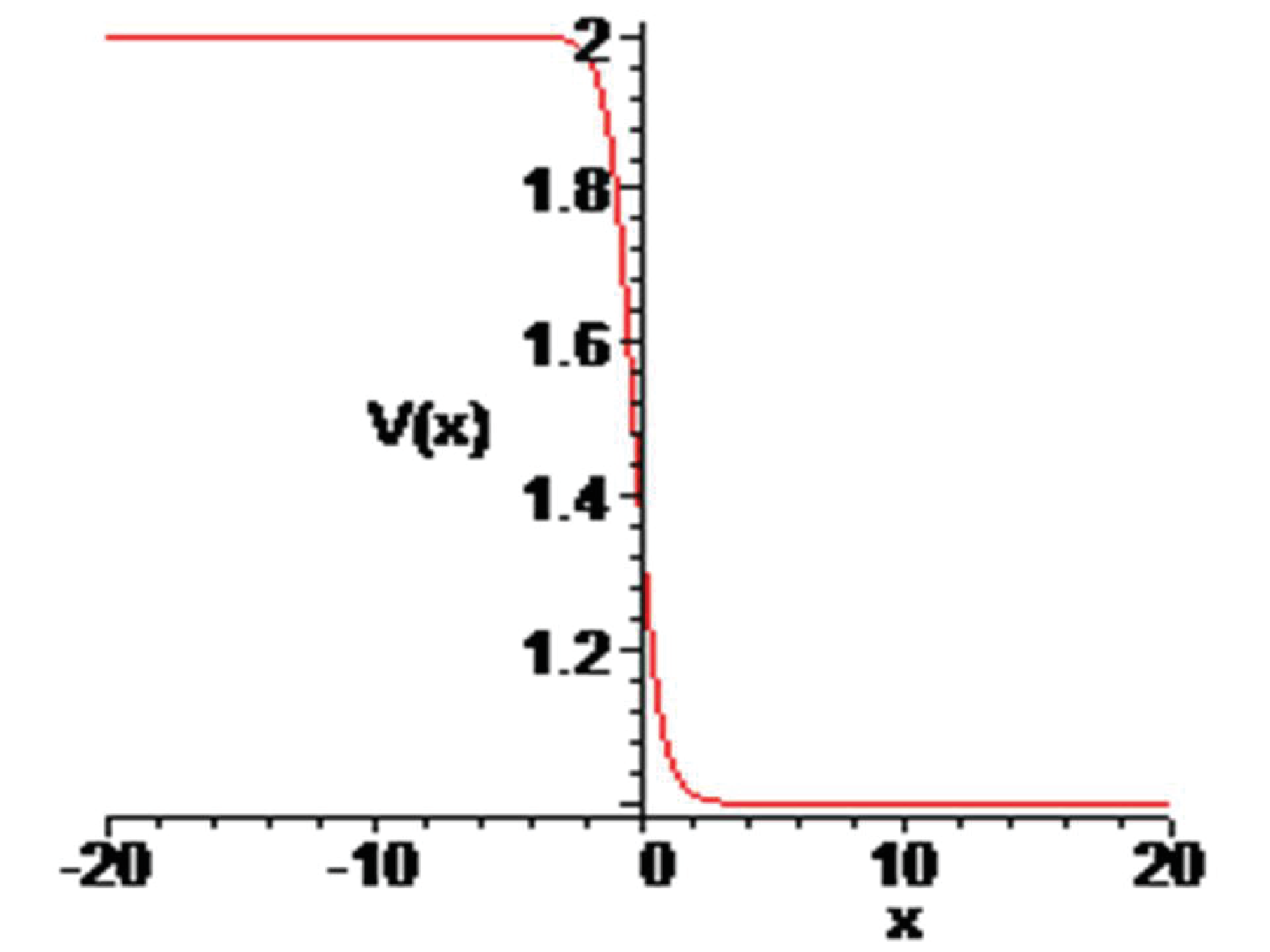,width=3in}}
\vspace*{1pt}
\caption{The Eckart potential with $V_{-\infty} = 2$, $V_{\infty} = 1$, $a = 1$, and $V_{0} = -1/9$.}\label{Ecpot}
\end{figure}

The Eckart potential takes the form
\begin{equation}
V(x) = \frac{V_{\infty} + V_{-\infty}}{2} + \frac{V_{\infty} - V_{-\infty}}{2}\tanh\left(\frac{x}{a}\right) + \frac{V_{0}}{\cosh^{2}(x/a)}.
\end{equation}
The shape of the Eckart potential is shown in Fig. \ref{Ecpot}. We define
\begin{equation}
k_{\pm\infty}^{2} = \frac{2m\left(E - V_{\pm\infty}\right)}{\hbar^{2}}~~\text{and}~~\bar{k} = \frac{k_{\infty} + k_{-\infty}}{2}.
\end{equation}

\begin{figure}[pb]
\centerline{\psfig{file=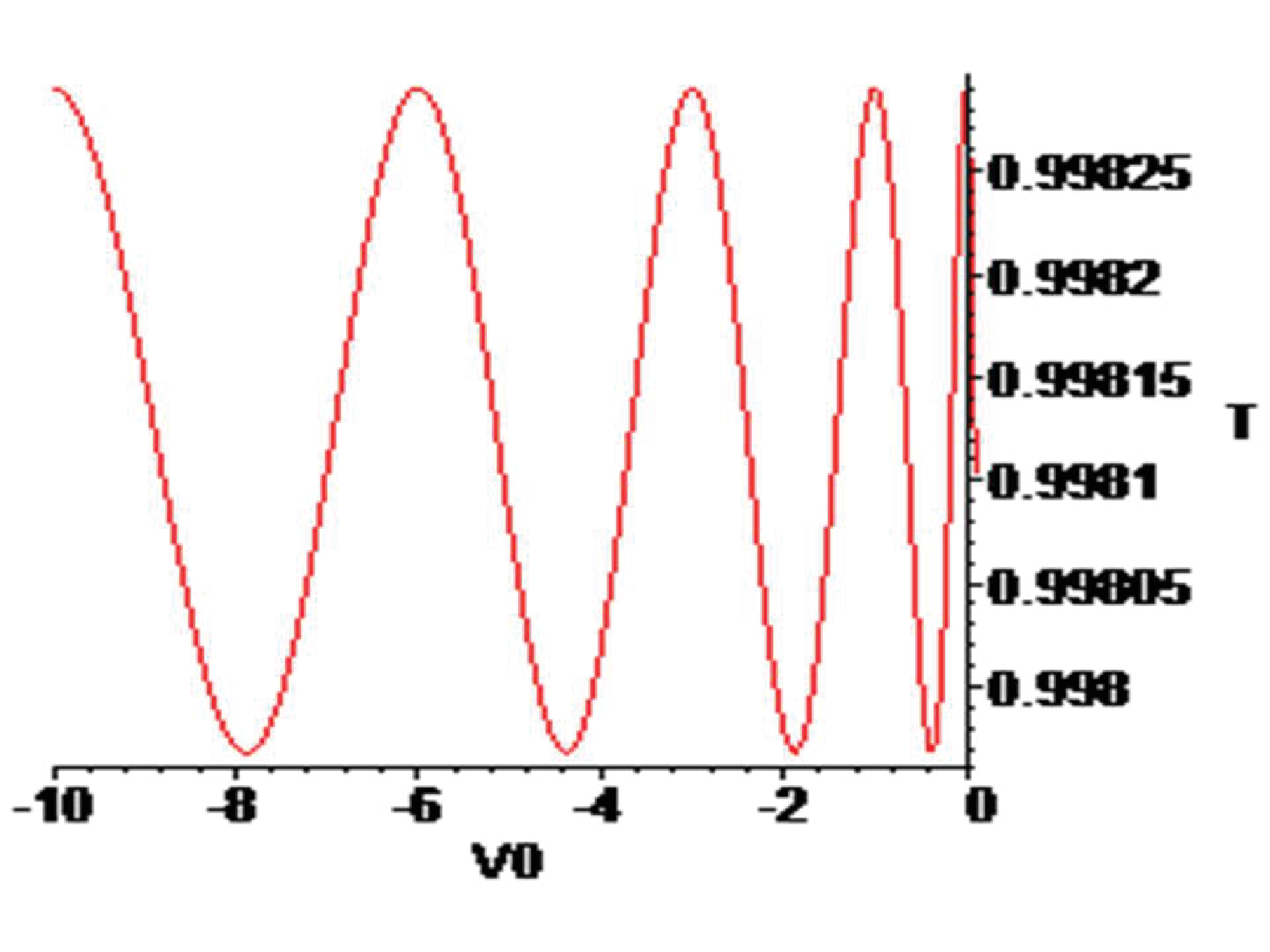,width=3in}}
\vspace*{1pt}
\caption{The transmission probabilities varying as $V_{0}$ in the Eckart potential $a = 1$, $k_{-\infty} = 1$, $k_{\infty} = 2$, and $m = \hbar = 1$.}\label{Tec}
\end{figure}

\begin{figure}[pb]
\centerline{\psfig{file=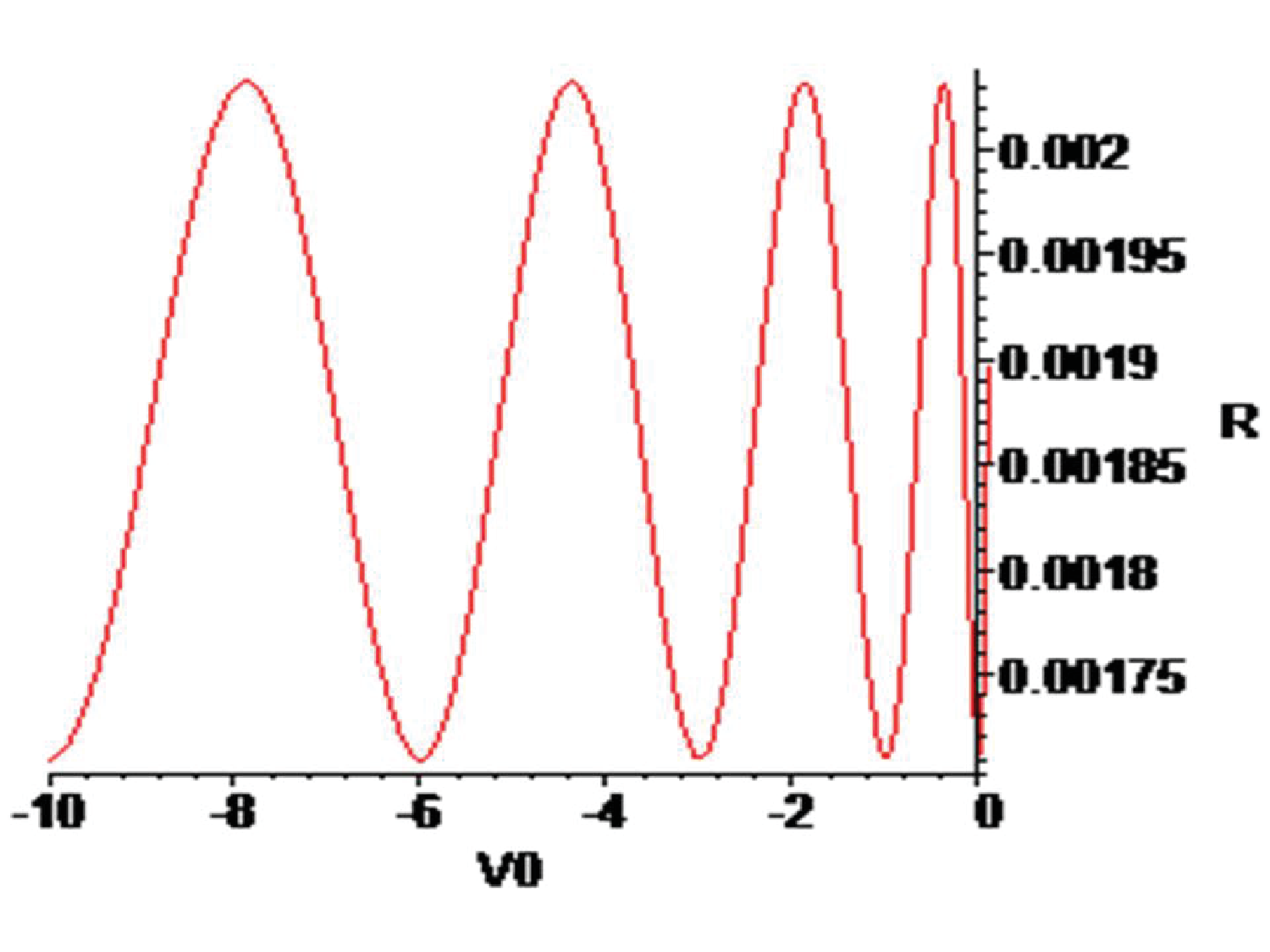,width=3in}}
\vspace*{1pt}
\caption{The reflection probabilities varying as $V_{0}$ in the Eckart potential with $a = 1$, $k_{-\infty} = 1$, $k_{\infty} = 2$, and $m = \hbar = 1$.}\label{Tec2}
\end{figure}

\noindent The transmission amplitude for this potential is given by (see \cite{Morse, Eckart})
\begin{eqnarray}
t &=& -\frac{i}{\sqrt{k_{\infty}k_{-\infty}}a}\frac{\Gamma[i\bar{k}a + (1/2) + \sqrt{(1/4) - 2mV_{0}a^{2}/\hbar^{2}}]}{\Gamma(ik_{\infty}a)}\nn\\
  &&  \times\frac{\Gamma[i\bar{k}a + (1/2) - \sqrt{(1/4) - 2mV_{0}a^{2}/\hbar^{2}}]}{\Gamma(ik_{-\infty}a)}.
\end{eqnarray}
A brief calculation shows that the transmission and reflection probabilities are
\begin{equation}
T = \frac{2\sinh(\pi k_{-\infty}a)\sinh(\pi k_{\infty}a)}{\cosh\left(2\pi\bar{k}a\right) + \cos\pi b}
\end{equation}
and
\begin{equation}
R = \frac{\cosh\left[\pi a\left(k_{-\infty} - k_{\infty}\right)\right] + \cos\pi b}{\cosh\left(2\pi\bar{k}a\right) + \cos\pi b},
\end{equation}
where $b = \sqrt{1 - 8mV_{0}a^{2}/\hbar^{2}}$. Fig. \ref{Tec} and \ref{Tec2} shows the transmission and reflection probabilities varying as $V_{0}$. From Fig. \ref{Tec}, we can see that the transmission resonances can occur. Analytically, the transmission resonances occur at
\begin{equation}
V_{0} = -\frac{\hbar^{2}}{2ma^{2}}n(n + 1).
\end{equation}
In contrast, there are no reflection resonances for this potential which can be seen from Fig. \ref{Tec2}.

\subsection{A (modified) Hulthen potential}
\begin{figure}[pb]
\centerline{\psfig{file=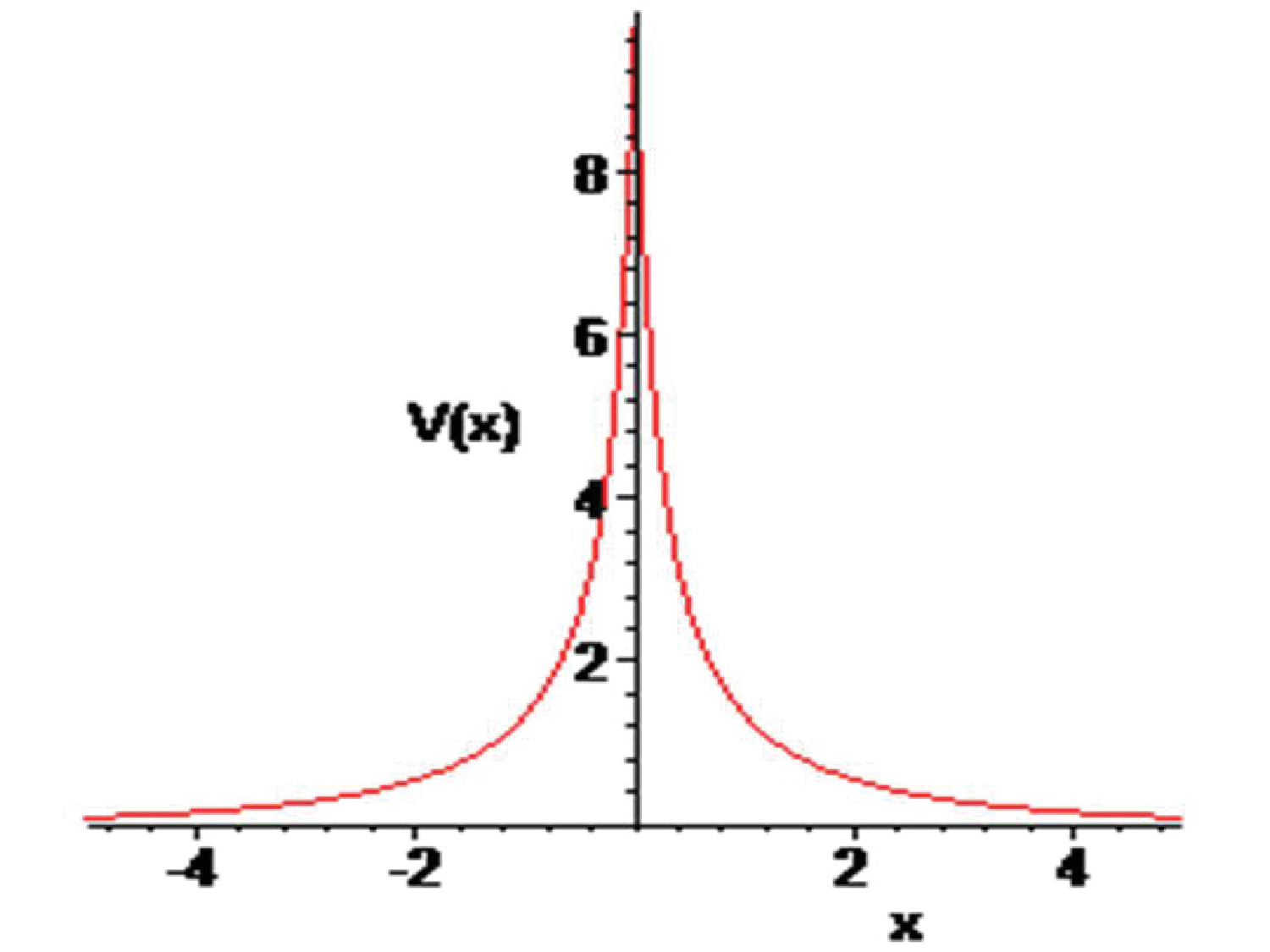,width=3in}}
\vspace*{1pt}
\caption{The Hulthen potential with $q = 0.9$, $a = 0.5$, and $V_{0} = 1$.}\label{Hulpot}
\end{figure}

A (modified) Hulthen potential takes the form \cite{Hulthen}
\begin{equation}
V(x) = \frac{V_{0}}{e^{a|x|} - q},
\end{equation}
where $V_{0}$, $a$, and $q$ all are real and positive with $q < 1$. The shape of the Hulthen potential is shown in Fig. \ref{Hulpot}. We are interested in two cases $E > V(x) > 0$ and $V(x) > E > 0$.

\subsubsection{Case I: $E > V(x) > 0$}
\begin{figure}[pb]
\centerline{\psfig{file=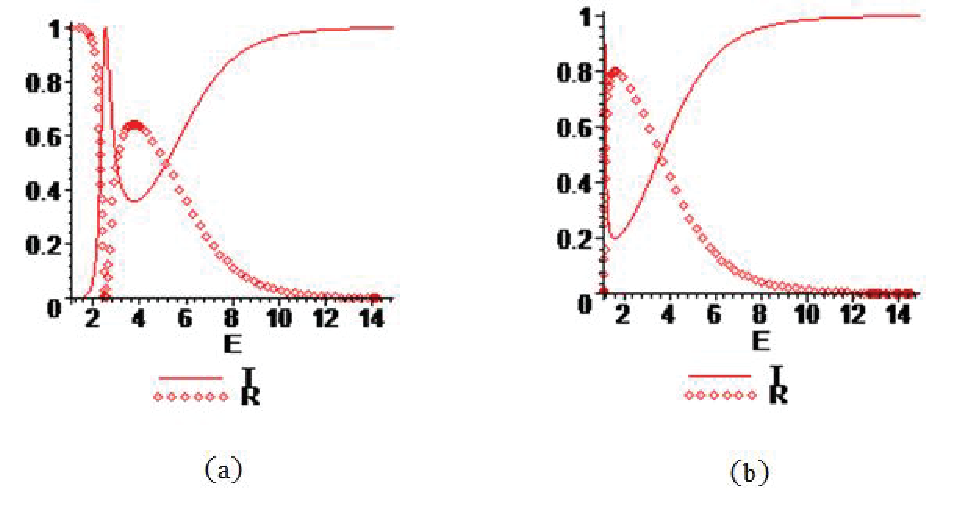,width=6in}}
\vspace*{1pt}
\caption{Plotting of transmission and reflection probabilities for varying $E$ for a Hulthen potential with (a) $a = 0.5$ and (b) $a = 1$ when $m = 1$, $V_{0} = 1$, and $q = 0.9$.}\label{Thul}
\end{figure}

The transmission and reflection amplitudes for this potential are given, in terms of a hypergeometric function, by (see \cite{Hulthen})
\begin{eqnarray}
t &=& \frac{(1 - q)^{2\lambda}q^{2\mu}}{1 + 2\mu}[\{q(1 + 2\mu)(\lambda^{2} - 2\lambda\mu + \mu^{2} - \nu^{2})F(1 + \lambda - \mu - \nu, 1 + \lambda - \mu + \nu, 2 - 2\mu; q)\nn\\
  && F(\lambda + \mu - \nu, \lambda + \mu + \nu, 1 + 2\mu; q)\} - \{q(1 - 2\mu)(\lambda^{2} + 2\lambda\mu + \mu^{2} - \nu^{2})\nn\\
  && F(1 + \lambda + \mu - \nu, 1 + \lambda + \mu + \nu, 2 + 2\mu; q)F(\lambda - \mu - \nu, \lambda - \mu + \nu, 1 - 2\mu; q)\}\nn\\
  && - \{(1 - 2\mu)(2\mu)(1 + 2\mu)F(\lambda + \mu - \nu, \lambda + \mu + \nu, 1 + 2\mu; q)\nn\\
  && F(\lambda - \mu - \nu, \lambda - \mu + \nu, 1 - 2\mu; q)\}]\nn\\
  && /[\{q(\lambda^{2} + 2\lambda\mu + \mu^{2} - \nu^{2})F(1 - \lambda - \mu - \nu, 1 - \lambda - \mu + \nu, 2 - 2\mu; q)\nn\\
  && F(\lambda - \mu - \nu, \lambda - \mu + \nu, 1 - 2\mu; q)\} + \{q(\lambda^{2} - 2\lambda\mu + \mu^{2} - \nu^{2})\nn\\
  && F(1 + \lambda - \mu - \nu, 1 + \lambda - \mu + \nu, 2 - 2\mu; q)F(-\lambda - \mu - \nu, -\lambda - \mu + \nu, 1 - 2\mu; q)\}\nn\\
  && - \{(2\mu)(1 - 2\mu)F(\lambda - \mu - \nu, \lambda - \mu + \nu, 1 - 2\mu; q)\nn\\
  && F(-\lambda - \mu - \nu, -\lambda - \mu + \nu, 1 - 2\mu; q)\}]
\end{eqnarray}
and
\begin{eqnarray}
r &=& -\frac{q^{1 + 2\mu}(\lambda^{2} + 2\lambda\mu + \mu^{2} - \nu^{2})}{1 + 2\mu}\sqrt{\frac{E + k}{E - k}}[\{(1 + 2\mu)F(\lambda + \mu - \nu, \lambda + \mu + \nu, 1 + 2\mu; q)\nn\\
  && F(1 - \lambda - \mu - \nu, 1 - \lambda - \mu + \nu, 2 - 2\mu; q)\} + \{(1 - 2\mu)\nn\\
  && F(1 + \lambda + \mu - \nu, 1 + \lambda + \mu + \nu, 2 + 2\mu; q)F(-\lambda - \mu - \nu, -\lambda - \mu + \nu, 1 - 2\mu; q)\}]\nn\\
  && /[\{q(\lambda^{2} + 2\lambda\mu + \mu^{2} - \nu^{2})F(1 - \lambda - \mu - \nu, 1 - \lambda - \mu + \nu, 2 - 2\mu; q)\nn\\
  && F(\lambda - \mu - \nu, \lambda - \mu + \nu, 1 - 2\mu; q)\} + \{q(\lambda^{2} - 2\lambda\mu + \mu^{2} - \nu^{2})\nn\\
  && F(1 + \lambda - \mu - \nu, 1 + \lambda - \mu + \nu, 2 - 2\mu; q)F(-\lambda - \mu - \nu, -\lambda - \mu + \nu, 1 - 2\mu; q)\}\nn\\
  && - \{(2\mu)(1 - 2\mu)F(\lambda - \mu - \nu, \lambda - \mu + \nu, 1 - 2\mu; q)\nn\\
  && F(-\lambda - \mu - \nu, -\lambda - \mu + \nu, 1 - 2\mu; q)\}],
\end{eqnarray}
where $\mu = ik/a$, $\nu = ip/a$, $\lambda = iV_{0}/aq$, $p^{2} = (E + V_{0}/q)^{2} - m^{2}$, and $k^{2} = E^{2} - m^{2}$. The transmission and reflection probabilities are derived from
\begin{equation}
T = |t|^{2}~~\text{and}~~R = |r|^{2}.
\end{equation}
We can check that
\begin{equation}
T + R = 1.
\end{equation}
The transmission and reflection probabilities for varying $E$ are shown in Fig. \ref{Thul}. There are both transmission and reflection resonances. Fig. \ref{Thul} (a) and \ref{Thul} (b) describe how the diffuseness $a$ has an effect on the probabilities with the other parameters $m$, $V_{0}$, and $q$ fixed.

\subsubsection{Case II: $V(x) > E > 0$}
\begin{figure}[pb]
\centerline{\psfig{file=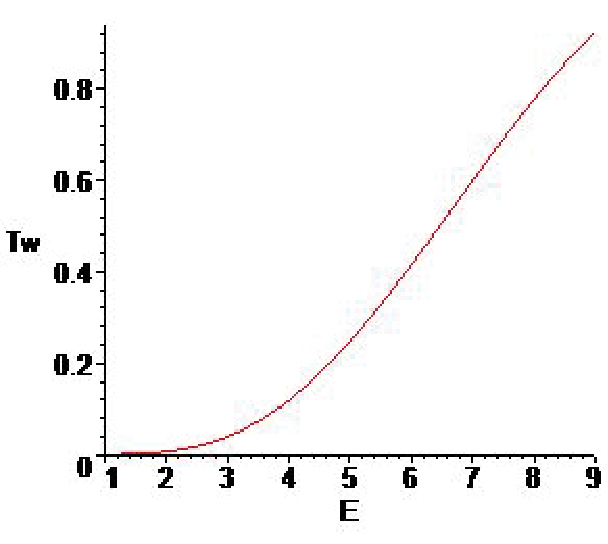,width=3in}}
\vspace*{1pt}
\caption{The transmission probability obtained from the WKB approximation with $V_{0} = 1$, $q = 0.9$, $a = 0.5$, and $m = \hbar = 1$.}\label{Thulwkb}
\end{figure}

By the WKB method, we obtain
\begin{equation}
T_{w} = \exp\left[-2\sqrt{\frac{2m}{\hbar^{2}}}\int_{-b}^{b}\sqrt{\frac{V_{0}}{e^{a|x|} - q} - E}\text{d}x\right],
\end{equation}
where $-b$ and $b$ are the classical turning points at which
\begin{equation}
\frac{V_{0}}{e^{ab} - q} - E = 0.
\end{equation}
That is
\begin{equation}
b = \frac{1}{a}\ln\left(\frac{V_{0}}{E} + q\right).
\end{equation}
Therefore,
\begin{equation}
T_{w} = \exp\left[-4\sqrt{\frac{2m}{\hbar^{2}}}\int_{0}^{(1/a)\ln(V_{0}/E + q)}\sqrt{\frac{V_{0}}{e^{ax} - q} - E}\text{d}x\right].
\end{equation}
To evaluate the integral we can proceed as follow: Let $z = e^{ax} - q$, so $\text{d}z = ae^{ax}\text{d}x = a(z + q)\text{d}x$. Then,
\begin{eqnarray}
T_{w} &=& \exp\left[-4\frac{\sqrt{2m}}{\hbar a}\int_{1 - q}^{V_{0}/E}\frac{1}{z + q}\sqrt{\frac{V_{0}}{z} - E}\text{d}z\right]\nn\\
      &=& \exp\left[-4\frac{\sqrt{2mV_{0}}}{\hbar a}\int_{1 - q}^{V_{0}/E}\frac{1}{z + q}\sqrt{\frac{1}{z} - \frac{E}{V_{0}}}\text{d}z\right].
\end{eqnarray}
Now let $u = 1/z$, so $\text{d}u = -(1/z^{2})\text{d}z = -u^{2}\text{d}z$. Then,
\begin{eqnarray}
T_{w} &=& \exp\left[4\frac{\sqrt{2mV_{0}}}{\hbar a}\int_{1/(1 - q)}^{E/V_{0}}\frac{1}{u + qu^{2}}\sqrt{u - \frac{E}{V_{0}}}\text{d}u\right]\nn\\
      &=& \exp\left[-4\frac{\sqrt{2mV_{0}}}{\hbar a}\int_{E/V_{0}}^{1/(1 - q)}\frac{1}{u(1 + qu)}\sqrt{u - \frac{E}{V_{0}}}\text{d}u\right].
\end{eqnarray}
Finally let $u - E/V_{0} = w^{2}$, so $\text{d}u = 2w\text{d}w$. Then,
\begin{eqnarray}
T_{w} &=& \exp\left[-8\frac{\sqrt{2mV_{0}}}{\hbar a}\int_{0}^{\sqrt{1/(1 - q) - E/V_{0}}}\frac{w^{2}}{(w^{2} + E/V_{0})(1 + qw^{2} + Eq/V_{0})}\text{d}w\right]\nn\\
      &=& \exp\left[-8\frac{\sqrt{2mV_{0}}}{\hbar aq}\int_{0}^{\sqrt{1/(1 - q) - E/V_{0}}}\frac{w^{2}}{(w^{2} + E/V_{0})(w^{2} + E/V_{0} + 1/q)}\text{d}w\right].
\end{eqnarray}
From
\begin{equation}
\int_{0}^{C}\frac{w^{2}}{(w^{2} + A)(w^{2} + B)}\text{d}w = \frac{1}{A - B}\left[\sqrt{A}\arctan\frac{C}{\sqrt{A}} - \sqrt{B}\arctan\frac{C}{\sqrt{B}}\right],
\end{equation}
we now obtain
\begin{equation}
T_{w} = \exp\left[-8\frac{\sqrt{2mV_{0}}}{\hbar a}\left[\sqrt{\frac{E}{V_{0}} + \frac{1}{q}}\arctan\sqrt{\frac{\dfrac{1}{1 - q} - \dfrac{E}{V_{0}}}{\dfrac{E}{V_{0}} + \dfrac{1}{q}}} - \sqrt{\frac{E}{V_{0}}}\arctan\sqrt{\frac{\dfrac{1}{1 - q} - \dfrac{E}{V_{0}}}{\dfrac{E}{V_{0}}}}\right]\right].
\end{equation}
Finally we see that in the WKB approximation
\begin{equation}
T_{w} = \exp\left[-8\frac{\sqrt{2mE}}{\hbar a}\left[\sqrt{1 + \frac{V_{0}}{Eq}}\arctan\sqrt{\frac{\dfrac{V_{0}}{(1 - q)E} - 1}{1 + \dfrac{V_{0}}{Eq}}} - \arctan\sqrt{\frac{V_{0}}{(1 - q)E} - 1}\right]\right].
\end{equation}
Fig. \ref{Thulwkb} shows the transmission probability for different heights of the Hulthen potential. The probability decreases if the height of the Hulthen potential increases.

\section{Conclusions}
In this paper, we used a $2 \times 2$ transfer matrix for a scattering process ($E > V(x)$) and the WKB method for a tunneling phenomenon ($E < V(x)$) \cite{1D, Bogoliubov, grey, mil, Analytic, Shabat, phd}. It is found that in some circumstances the WKB method is highly accurate when compared to exact solutions. We could use this WKB method to find reflection and transmission probabilities for difficult potentials which may give no exact solutions.

\section*{Acknowledgement}
This research was supported by a grant for the professional development of new academic staff from the Ratchadapisek Somphot Fund at Chulalongkorn University, by Thailand Toray Science Foundation (TTSF), and by the Research Strategic plan program (A1B1), Faculty of Science, Chulalongkorn University. PB was additionally supported by a scholarship from the Royal Government of Thailand. TN was additionally supported by a scholarship from the Development and Promotion of Science and Technology talent project (DPST). TN gives a special thanks to Dr. Auttakit Chatrabuti for his valuable advice. We also thank Prof. Matt Visser for extremely useful suggestions and comments.


\begin{thebibliography}{}
\bibitem{Landshoff}
P. V. Landshoff and A. Metherell, Simple Quantum Physics. University Press, Cambridge: 1997.

\bibitem{Landau}
L. D. Landau and E. M. Lifshitz, Quantum Mechanics: Non-Relativistic Theory. Pergamon, New York: 1977.

\bibitem{Baym}
G. Baym, Lectures on Quantum Mechanics. Benjamin, New York: 1969.

\bibitem{Gasiorowicz}
S. Gasiorowicz, Quantum Physics. Wiley, New York: 1996.

\bibitem{Capri}
A. Z. Capri, Non-Relativistic Quantum Mechanics. Benjamin-Cummings, Menlo Park, California: 1985, pp. 95-109.

\bibitem{Stehle}
P. Stehle, Quantum Mechanics. Holden-Day, San Francisco: 1996 pp. 57-60.

\bibitem{Schiff}
L. I. Schiff, Quantum Mechanics. McGraw-Hill, New York: 1955.

\bibitem{Cohen}
C. Cohen-Tannoudji, B. Dui, and F. Lal\"{o}e, Quantum Mechanics. Wiley, New York: 1977.

\bibitem{Galindo}
A. Galindo and P. Pascual, Quantum Mechanics I. Springer-Verlag, Berlin: 1990.

\bibitem{Park}
D. Park, Introduction to the Quantum Theory. McGraw-Hill, New York: 1974.

\bibitem{Fromhold}
A. T. Fromhold, Quantum Mechanics for Applied Physics and Engineering. Academic, New York: 1981.

\bibitem{Scharff}
M. Scharff, Elementary Quantum Mechanics. Wiley, London: 1969.

\bibitem{Messiah}
A. Messiah, Quantum Mechanics. North-Holland, Amsterdam: 1958.

\bibitem{Merzbacher}
E. Merzbacher, Quantum Mechanics. Wiley, New York: 1965.

\bibitem{Singh}
J. Singh, Quantum Mechanics: Fundamentals and Applications to Technology. Wiley, New York: 1997.

\bibitem{Mathews}
P. M. Mathews and K. Venkatesan, A Textbook of Quantum Mechanics. McGraw-Hill, New York: 1978.

\bibitem{1D}
M. Visser, ``Some general bounds for 1-D scattering," Phys. Rev. A 59, 1999 pp. 427-438, [arXiv: quant-ph/9901030].

\bibitem{Bogoliubov}
P. Boonserm and M. Visser, ``Bounding the Bogoliubov coefficients," Annals of Physics 323, 2008, 27792798 [arXiv: quant-ph/0801.0610].

\bibitem{grey}
P. Boonserm and M. Visser, ``Bounding the greybody factors for Schwarzschild black holes," Phys. Rev. D 78, 2008, 101502, [Rapid Communications], [arXiv:gr-qc/0806.2209].

\bibitem{mil}
P. Boonserm and M. Visser, ``Transmission probabilities and the Miller-Good transformation," Journal of Physics A: Mathematical and Theoretical 42, 2009, 045301, [arXiv: math-ph/0808.2516].

\bibitem{Analytic}
P. Boonserm and M. Visser, ``Analytic bounds on transmission probabilities," Ann. Phys. 325, 2010, 1328-1339 [arXiv:0901.0944 [math-ph]].

\bibitem{Shabat}
P. Boonserm and M. Visser, ``Reformulating the Schr\"{o}dinger equation as a Shabat-Zakharov system," Journal of Mathematical Physics 51, 2010, 022105, [arXiv:0910.2600 [math-ph]].

\bibitem{phd}
P. Boonserm, Rigorous Bounds on Transmission, Reflection, and Bogoliubov Coefficients. (PhD thesis), [arXiv:0907.0045 [math-ph]].

\bibitem{Ushveridze}
A. G. Ushveridze, Quasi-Exactly Solvable Models in Quantum Mechanics. Institute of Physics Publishing, London: The Institute of Physics: 1994.

\bibitem{Solid}
A. van der Ziel, Solid State Physical Electronics. Englewood Cliffs, NJ, Prentice Hall: 1976.

\bibitem{Brandsen}
B. H. Brandsen and C. J. Joachain, Quantum Mechanics. Prentice Hall, New York: 2000.

\bibitem{Morse}
P. M. Morse and H. Feshbach, Methods of Theorectical Physics. McGraw-Hill, New York: 1953.

\bibitem{Eckart}
C. Eckart, ``The penetration of a potential barrier by electrons", Phys. Rev. 35, 1930, p. 1303.

\bibitem{Hulthen}
J. Y. Guo, Y. Yu, and S. W. Jin, ``Transmission resonance for a Dirac particle in a one-dimensional Hulthen potential", Eur. J. Phys. 7, 2009, pp. 168-174.
\end{thebibliography}
\end{document}